\documentclass[twocolumn,prb, showpacs]{revtex4-1}
\usepackage{graphicx}
\usepackage{dcolumn}
\usepackage{bm}
\usepackage{color}

\begin{document}


\title{$\pi$-bonds in graphene and surface of topological insulator $Bi_2Se_3$}

\author{G. J. Shu$^1$}
\author{S. C. Liou$^1$}
\author{F. C. Chou$^{1,2}$}
\email{fcchou@ntu.edu.tw}
\affiliation{
$^1$Center for Condensed Matter Sciences, National Taiwan University, Taipei 10617, Taiwan}
\affiliation{
$^2$National Synchrotron Radiation Research Center, Hsinchu 30076, Taiwan}

\date{\today}

\begin{abstract}
Chemical bond analysis for the topological insulator (TI) Bi$_2$Se$_3$ following valence bond theory and VSEPR rule is proposed and compared to that of graphene. The existence of $\pi$-bond trimers on surface Se-layer is revealed to be responsible for the unusual 2D surface conduction found in the prototype $Z_2$ topological insulator Bi$_2$Se$_3$, which shows great similarity to the ideal ``surface-only" 2D material graphene with surface band in Dirac cone shape.  The $\pi$-bond trimers on the surface of TI form a dynamic $\pi$-bond conjugated system of implicit chirality.  Local $\pi$-bond energy exchange in cross-bridge model is proposed responsible for the unique 2D surface conduction for topological insulator Bi$_2$Se$_3$, and the similar low dimensional conduction mechanism can also be identified in graphene and conductive polymers.  Preliminary supporting evidence for the existence of $\pi$-bonds is provided by the commonly assigned $\pi$ plasmon for the spectral feature near 7 eV in graphite and Bi$_2$Se$_3$ using electron energy-loss spectroscopy (EELS).      
\end{abstract}


\maketitle

The finding of topological insulator has been one important breakthrough in condensed matter physics since Kane and Mele first suggested that the Dirac cone band structure of graphene can also be found in this novel class of material.\cite{Kane2005}  Two major classes of TI have been predicted from calculation and verified experimentally by ARPES since; the first type has surface state protected by the time reversal symmetry (TRS) called $Z_2$ topological insulators, such as Bi$_2$Se$_3$,\cite{Hasan2010} and the second type is the topological crystalline insulators with surface state protected by the crystalline symmetry, such as Pb$_{1-x}$Sn$_x$Se.\cite{Fu2007, Dziawa2012}  Simply put, topological insulator is a unique quantum state of matter of narrow band gap with surface-only conduction as a result of strong spin-orbit coupling, and, in particular, the $Z_2$ TI has a surface spin-polarized state protected by the TRS topologically.  In the language of theoretical calculations commonly used in physics, TI is expected to be found in a material system which has opposite parity for the conduction and valence bands, and the occurrence of band inversion is triggered by parameters such as strain or strong spin-orbit coupling (SOC).\cite{Zhang2009, Fu2007}    However, the Hilbert space topology often used by theorists is too abstract for chemists and material scientists for further material design.\cite{Hasan2010, Qi2011}  We believe that a real space view from the chemical bond perspective should enhance and be complimentary to the understanding and material design of TI.  In particular, with the help of the supporting experimental evidence using EELS technique, we find that there is a great deal of similarity between the ideal 2D material of ``surface-only" graphene and Bi$_2$Se$_3$ surface in terms of the Dirac cone band picture and the existence of $\pi$-bonds. 

Molecular Orbital (MO) theory has shown great success in describing condensed matter in the energy-momentum space, i.e, the lowest-unoccupied-molecular-orbital (LUMO) and the highest-occupied-molecular-orbital (HOMO) for individual molecule could be extended to the valence and conduction bands of solids.\cite{textbook}  However, except for the typical metals such as pure IA and IIA elements, insulators of ionic nature, and the wide band gap semiconductors, more ``exceptions" are found in the band theory of solids.  For example, Coulomb repulsion $U$ is introduced to the strongly correlated electron system that has an unfilled band signified to be metallic but turns out to be insulating, and often categorized as Mott insulator.\cite{Mott}  The latest development of topological insulator brings up yet another ``exception" for a narrow band gap semiconductor, i.e., a band insulator for the bulk with a spin polarized conducting surface state, which can hardly be defined as either a zero gap semiconductor or a conductor with zero density of states in the conduction band.

The unique $sp^2$+$p$ orbital hybridization of graphene allows the honeycomb structure formation at the lowest 2D atomic packing density, and one electron remains in $p_z$-orbital.\cite{graphene}  It is desirable to find the link between the real space chemical bond of $sp^2+p$ and the unique energy band of gapless Dirac cone shape.  While the honeycomb lattice is bridged by the covalent $\sigma$-bonds through $sp^2$ hybridization, the remaining electron in $p_z$ is unpaired with relatively higher energy without the $\sigma$-bond energy gain.  Judging from the metallic nature of graphene, the $p_z$ electrons cannot be localized completely, yet the traditional band picture cannot describe the co-existing bonding and nonbonding states reasonably well. Such a dilemma could be resolved by the unique bond order of 1$\frac{1}{3}$ for graphene, i.e., one $\pi$-bond is shared by three neighboring caron atoms statistically.\cite{Fasolino2007}  In fact, the linear dispersion relationship of energy-momentum suggests the tunneling nature of a massless fermion excitation.\cite{Geim2009, Katsnelson2006}  The conduction mechanism shown in Dirac cone shape band picture must be closely related to the semi-localized $p_z$ electron behavior in graphene.  While the topological insulator Bi$_2$Se$_3$ shows Dirac cone surface band similar to that of graphene, and with the reasonable assumption on the close connection between the dynamics of the unpaired $p_z$ electrons and the Dirac cone surface band picture, it would be important to find out whether there also exist unpaired electrons of similar dynamic nature on the surface of Bi$_2$Se$_3$.

\begin{figure}
\begin{center}
\includegraphics[width=3.5in]{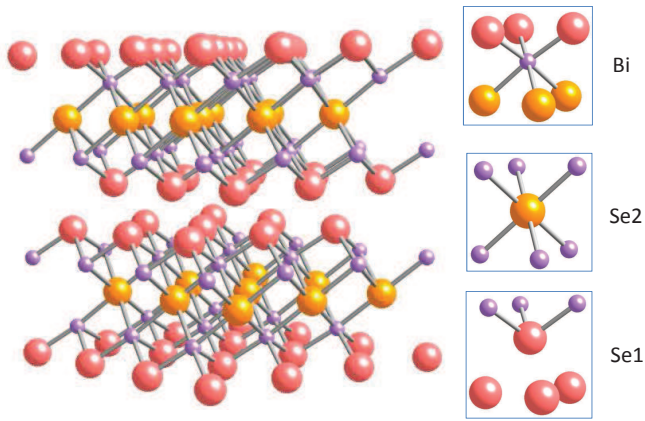}
\end{center}
\caption{\label{fig-structure}(color online) The crystal structure of Bi$_2$Se$_3$ is shown with two quintuple layers. The insets show the actual octahedral environment near Bi, Se2 in the middle, and Se1 at the van der Waals gap. }
\end{figure}

\begin{figure}
\begin{center}
\includegraphics[width=3.5in]{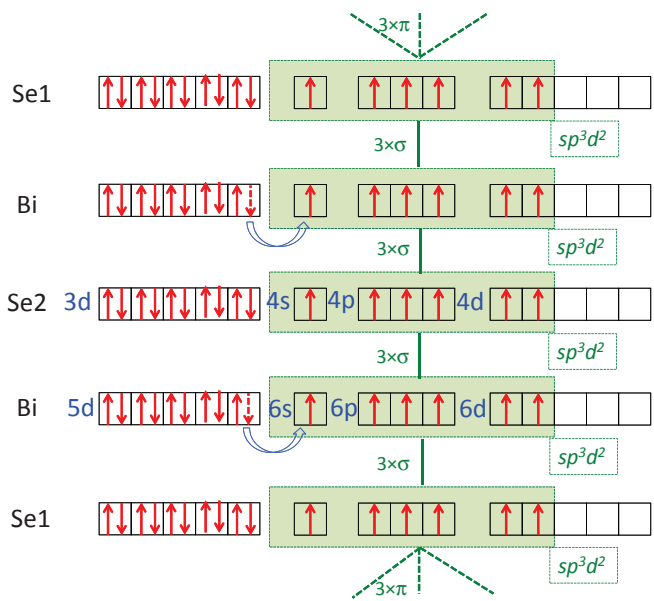}
\end{center}
\caption{\label{fig-VBmodel}(color online) The valence bond model of Bi$_2$Se$_3$ constructed from atomic electronic configurations of Bi ([Xe]5d$^{10}$6s$^2$6p$^3$) and Se ([Ar]4s$^2$4p$^4$) through $sp^3d^2$ hybridization for the octahedral coordination, which is adapted from the VB model for Bi$_2$Te$_3$ proposed by Drabble \textit{et al.}\cite{Drabble1958} }
\end{figure}

Single Dirac cone at the $\bar{\Gamma}$-point has been predicted and experimentally confirmed in the $Z_2$ topological insulator Bi$_2$Se$_3$.\cite{Hasan2010}  The real space view of the surface chemical bond of Bi$_2$Se$_3$ should be examined in order to find out the peculiar non-localization of surface electrons that maybe related to the existence of Dirac cone.  While Bi$_2$e$_3$ is a narrow band gap semiconductor of E$_g$$\sim$0.3 eV, a complete $\sigma$-bond construction following valence bond (VB) theory should be applicable using orbital hybridization, similar to that of the $sp^3$ hybridization for the typical semiconductor silicone with E$_g$$\sim$1 eV.   The crystal structure of Bi$_2$Se$_3$ can be described as layers of quintuple unit with van der Waals interlayer coupling, as shown in Fig.~\ref{fig-structure}.\cite{Huang2012}  The chemical bonds of Bi$_2$Te$_3$ with identical crystal structure has been under debate since Mooser and Pearson first introduced the concepts of orbital hybridization and resonating valence bonds to the system.\cite{Mooser1958, Drabble1958, Bhide1971}  The focus of the early debate was on what kind of hybrid orbitals for Bi and Te are the most suitable for explaining the observed transport property well.  However, the early proposed hybridization models have not taken the van der Waals and surface chemical bondings into consideration.  The quintuple unit of Bi$_2$Se$_3$ shown in Fig.~\ref{fig-structure} indicates that all of the Bi and Se atoms have octahedral environment, which implies that the $sp^3d^2$ hybrid is required under the valence bond theory and VSEPR rule consideration.\cite{inorganic}  Fig.~\ref{fig-VBmodel} shows the VB model proposed for Bi$_2$Se$_3$ starting from the atomic electronic configuration.  We find that the $sp^3d^2$ hybrid is likely for all Se atoms because of the required six outer shell electrons in n=4; on the other hand, it is required to move one electron from 3d$^{10}$ to satisfy the required $sp^3d^2$ hybrid for the two Bi atoms.  The heavy elements of both Bi and Se suggests that the formed covalent $\sigma$-bonds must be relatively weak considering the large atom size of high principal quantum number n ($>$4), and indeed, the melting point of Bi$_2$Se$_3$ is relatively low near 705$^\circ$C.  As a typical layered compound with van der Waals force in between quintuple layers, it is interesting to note that the Se atoms near the van der Waals gap have three unpaired electrons remaining at the gap interface, and these unpaired electrons will be exposed to the surface when the sample is cleaved open along the van der Waals gap.  

\begin{figure}
\begin{center}
\includegraphics[width=3in]{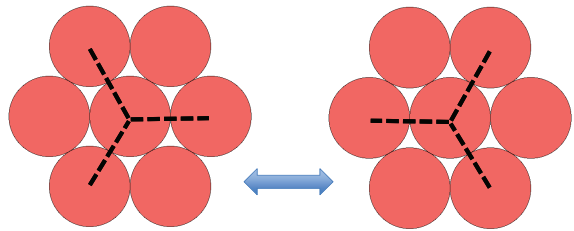}
\end{center}
\caption{\label{fig-trimer}(color online) The $\pi$-bond trimers on surface Se-layer of coordination number six in Bi$_2$Se$_3$ form a $\pi$-bond conjugated system (in dashed lines) with an implicit chirality. }
\end{figure}

Usually, the unpaired electrons exposed to the cleaved surface could end up stabilizing into different surface states, including passivation by the external molecules, surface phonon-related reconstruction that leads to the band bending, or collective itinerant behavior like surface plasmon.  We believe that the most interesting and critical condition is when the unpaired electrons carry energy that is not too high to be itinerant or too low to be localized, as often seen in the low melting point semimetal materials with narrow band gap - Bi$_2$Se$_3$ and most confirmed topological insulators seem to fall in this category.\cite{TI}  In view of the weak bonding energy of Bi$_2$Se$_3$ and the surface hexagonal close packing of Se atoms in vacuum, we believe that the three unpaired electrons per Se on the exposed surface could lead to a novel surface state that is related to the unique surface property for the topological insulators.  We propose that the three unpaired electrons in virtual octahedron in $sp^3d^2$ hybrid per Se1 could form a critical state as trimer of 120$^\circ$ apart to avoid the Coulomb repulsion, in accord with the VSEPR rule, as shown in Fig.~\ref{fig-trimer}.  Since all Se1 atoms are in 2D triangular lattice of hexagonal close packing on the surface, the three unpaired electrons per surface Se1 must either avoid each other to reduce Coulomb repulsion, or form $\pi$-bond with the three neighboring surface Se1 atoms instantaneously.  It should be noted that when the proposed $\pi$-bond is weak with minimum orbital overlap on the surface, the reduction of Gibbs free energy through entropy gain ($-T\Delta S$) is not negligible anymore, i.e., the $\pi$-bond trimer could form a dynamic conjugated system instead of stabilizing in one of the configurations thermodynamically, as illustrated in Fig.~\ref{fig-trimer}.  In fact, such a $\pi$-bond conjugated system is commonly found in graphene, benzene, superconductor MgB$_2$, and conductive polymers.\cite{Choi2002, ppp, Fasolino2007}

$\pi$-bond is a unique class of chemical bond formed with instantaneous Coulomb attraction between semi-localized electrons in mostly side-to-side orbitals without actual overlap, in contrast to the $\sigma$-bond which requires electron sharing in the head-to-head overlapped orbital space.\cite{inorganic}  Because of its extremely weak bonding strength and the dynamic conjugated nature which is easily perturbed by thermal or strain energy, it is difficult to describe $\pi$-bond (or $\pi$-band) in energy-momentum space for the ground state description.  In fact, while graphene is in the thickness limit of one molecular thick perpendicular to its 2D honeycomb lattice, it is not appropriate to describe the related physical phenomena in a continuous band picture anymore; instead, a discrete quantum state for electron in $p_z$ orbital with the greatest uncertainty on momentum must exist, according to the Uncertainty Principal.  Quantum tunneling phenomenon of Klein paradox and the Dirac equation for massless fermions have all pointed toward this direction for the description of electrons near the Dirac cone for graphene.\cite{Katsnelson2006}  Following the link between the conjugated $\pi$-bond system and the existence of Dirac cone for graphene, the commonly observed gapless Dirac cone on topological insulator surface suggests that the similar $\pi$-bond conjugated system could also be identified in Bi$_2$Se$_3$.  

\begin{figure}
\includegraphics[width=3.5in]{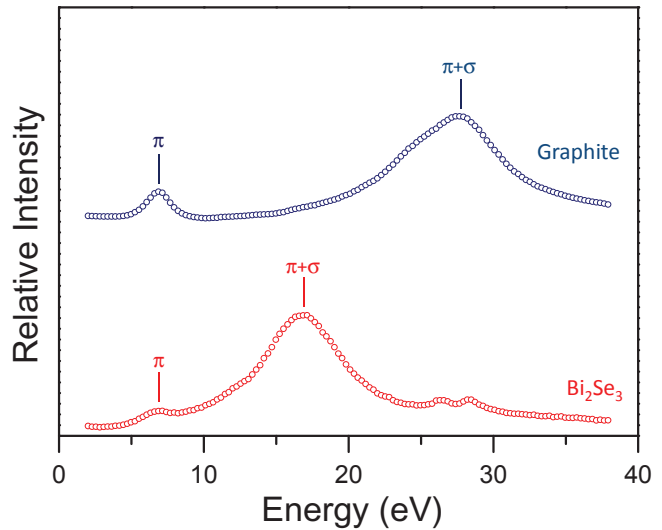}
\caption{\label{fig-EELS}(color online) The STEM-EELS spectra for graphite and Bi$_2$Se$_3$ are compared.  The first two main features are assigned coming from the $\pi$ and $\pi+\sigma$ plasmons as described in the text. } 
\vspace{-5mm}
\end{figure}

Herein, we offer a positive experimental evidence to the existence of $\pi$-bond using electron energy-loss spectroscopy (EELS) through comparative studies of graphite (graphene) and topological insulator Bi$_2$Se$_3$.  EELS can provide the information of both valence and conduction band electronic states effectively.  In particular, the low-loss region (energy loss $<$ 50 eV) in EELS spectra contains useful information on the collective modes of valence electron excitation, e.g., surface and volume plasmons; and the single-particle excitations, e.g., interband transitions and low-lying core-level ionizations.\cite{Raether1980, Egerton2011}  The representative EELS spectra for graphite and Bi$_2$Se$_3$ are compared in Fig.~\ref{fig-EELS}.\cite{Liou2013}   The spectral features near 7 and 27 eV for graphite have been demonstrated as the plasma resonances related to the $\pi$ and $\pi$+$\sigma$ electrons, respectively.\cite{Ichikawa1958, Taft1965}  The effective number of electrons per atom ($n_{eff}(\omega)$) participating in the collective oscillation modes near 7 eV and 27 eV have been estimated to be $n_{eff}$$\sim$1 and 4 electrons, respectively.\cite{Taft1965}  The extracted $n_{eff}$ values strongly suggest that the participating electrons for the assigned $\pi$ plasmon (one $\pi$-bond) and the $\pi$+$\sigma$ plasmon (one $\pi$- plus three $\sigma$-bonds) are coming from the unpaired electron in $p_z$ orbital and the three electrons in the hybridized $sp^2$ orbitals per atom, respectively.   

The EELS spectrum for Bi$_2$Se$_3$ shown in Fig.~\ref{fig-EELS} reveals two main spectral features near 7 and 17 eV, which are in good agreement with the spectra reported in the literature.\cite{Nascimento1999}   The two weak peaks at $\sim$26.4 and 28.4 eV can be identified as the Bi \textit{O}$_{4,5}$-edge excitation from Bi $5d$ electrons, which agrees with the $5d^9$ configuration after the required $sp^3d^2$ hybridization shown in Fig.~\ref{fig-VBmodel}.\cite{Liou2013, Nascimento1999}  We can tentatively assign the two main spectral features near 7 and 17 eV to be volume plasmons first.  Considering there are six $\pi$-bonds and twelve $\sigma$-bonds per quintuple Bi$_2$Se$_3$ formula unit (see Fig.~\ref{fig-VBmodel}), we expect the participating $n_{eff}$ for the $\pi$ and $\pi+\sigma$ plasmons for Bi$_2$Se$_3$ be close to 1.2 (6/5 per atom, estimated from the six $\pi$-bonds per quintuple unit) and 3.6 (18/5 per atom, estimated from the 6$\pi$+12$\sigma$ bonds per quintuple unit), respectively.  The $n_{eff}$ derived from the energy-loss function near 7 and 17 eV are $\sim$1 and $\sim$3 electrons per atom, which can be largely interpreted as coming from the plasma oscillations of the $\pi$ and $\pi + \sigma$ electrons, respectively.\cite{Liou2013}  The reasonable agreement between the $\pi$-bond model prediction and the EELS estimation strongly supports the existence of $\pi$-bond in Bi$_2$Se$_3$. 

\begin{figure}
\begin{center}
\includegraphics[width=3.5in]{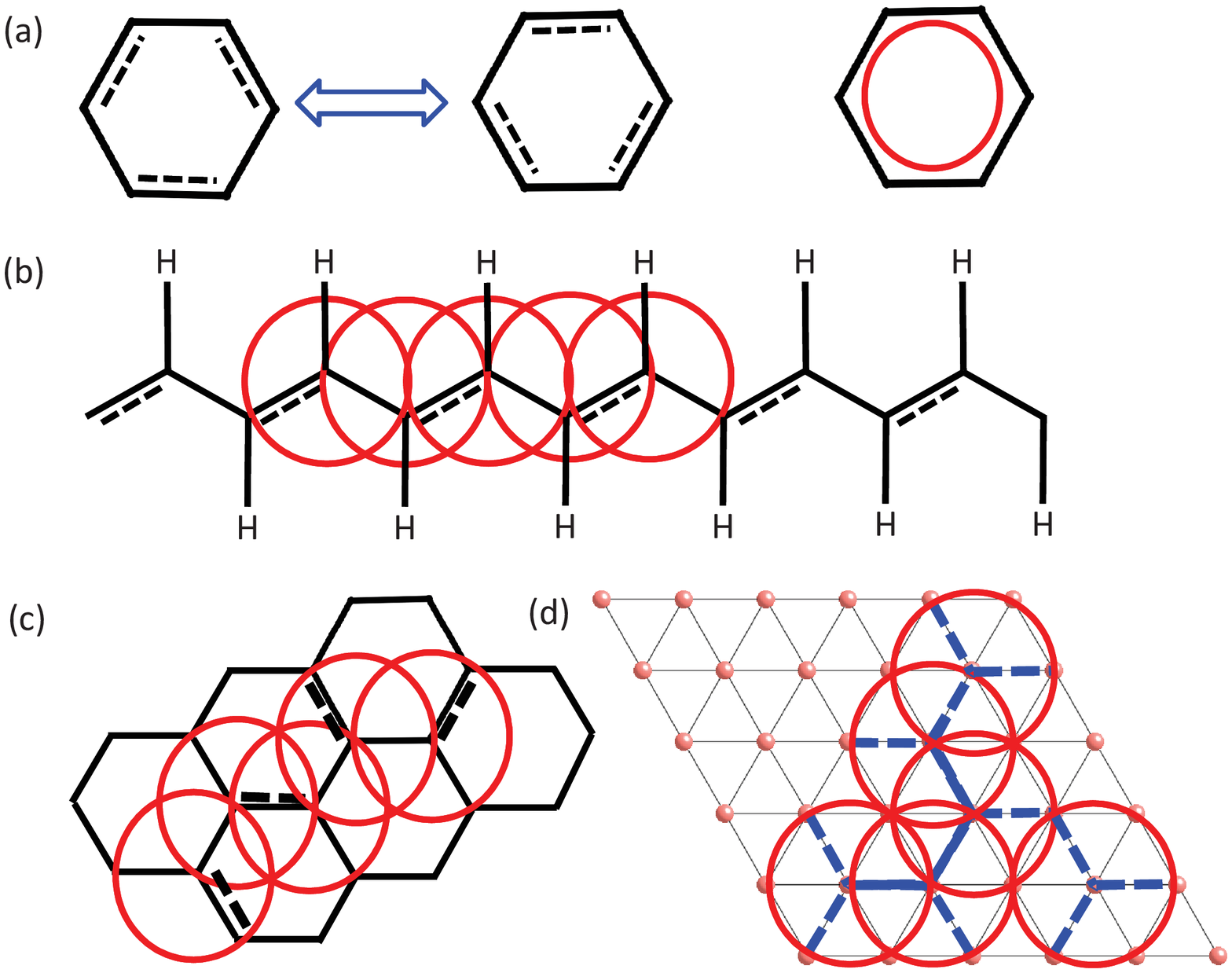}
\end{center}
\caption{\label{fig-piBonds}(color online) The $\pi$-bonds (dashed line) can be identified in (a) benzene C$_6$H$_6$ with closed $\pi$-bond conjugated system, (b) trans-polyacetylene [C$_2$H$_2$]$_n$ with 1D open $\pi$-bond conjugated system, (c) graphene with 2D open $\pi$-bond conjugated system, and (d) Bi$_2$Se$_3$ with 2D $\pi$-bond trimer conjugated system. The red circle represents the dynamic $\pi$-bond conjugated system of inherited chirality.}
\end{figure}

The $\pi$-bond conjugated system has been applied to successfully explain the unique transport behavior of conductive polymer.\cite{ppp}  As shown in Fig.~\ref{fig-piBonds}(b), the 1D conduction for trans-polyacetylene [C$_2$H$_2$]$_n$ can be explained by the local energy transfer through the instantaneous $\pi$-bond bonding/nonbonding of the $\pi$-bond conjugated system.  In contrast, benzene has three $\pi$-bonds within each unit (Fig.~\ref{fig-piBonds}(a)), but the formed conjugated system is closed, which leads to its insulating nature.  It is interesting to note that the 1D bridged $\pi$-bond conjugated system shown in trans-polyacetylene [C$_2$H$_2$]$_n$ can be extended to 2D and form one honeycomb network similar to graphene, as illustrated in Fig.~\ref{fig-piBonds}(c).  We believe that the 2D transport behavior of graphene can be explained using the same open $\pi$-bond conjugated system.  Considering the $\pi$-bond trimer conjugated system found on the Bi$_2$Se$_3$ surface, similar transport mechanism can also apply, as illustrated in Fig.~\ref{fig-piBonds}(d).  As long as all surface Se1 atoms contain three unpaired electrons of energy not too high to be itinerant or too low to be localized, the critical energy equivalent to the $\pi$-bond bonding energy would co-exist with the entropy-dominated dynamic $\pi$-bond conjugated system, which can also explain why TI is so sensitive to strain as found in  the theoretical prediction.\cite{Fu2007}  The $\pi$-bond conjugated trimer could bond and break spontaneously (i.e., with nonzero configurational entropy) and the small critical bonding energy is transferred locally, which would lead to a successful long range 2D electronic energy transfer beyond the traditional concept of electric conduction based on diffusion of itinerant electrons.  We might name such 2D conduction mechanism as a ``cross-bridge" model, because the tri-bridge with inherited chirality of random handedness can always maintain an infinite percolation route in 2D at any instance.

In summary, with the help of chemical bond analysis under valence bond theory and VSEPR rule considerations, $sp^3d^2$ hybridization for Bi and Se atoms in Bi$_2$Se$_3$ leads to the revelation of $\pi$-bond trimer on the surface of topological insulator Bi$_2$Se$_3$.  The existence of $\pi$-bond in both graphite and Bi$_2$Se$_3$ has been supported from EELS, which not only helps to explain the unique 2D conduction mechanism in the ``cross-bridge" model , but also helps to explain the unique spin polarization of Dirac cone-shaped surface band for topological insulator.  Current real space description of Bi$_2$Se$_3$ can be valuable to understand the topological insulator beyond the reciprocal space view.  The role of $\pi$-bond on topological insulator surface could be the key to the understanding of both graphene and topological insulators in general.       

\section*{Acknowledgment}
We thank G. Y. Guo, M.-W. Chu, and C. H. Chen for helpful discussions and R. Sankar for the single crystal sample Bi$_2$Se$_3$ used in the EELS studies.  FCC acknowledges support from NSC-Taiwan under project number  NSC 101-2119-M-002-007.  GJS acknowledges support from NSC-Taiwan under project number NSC 100-2112-M-002-001-MY3.




\end{document}